\documentclass[manuscript,screen]{acmart}

\AtBeginDocument{%
  \providecommand\BibTeX{{%
    \normalfont B\kern-0.5em{\scshape i\kern-0.25em b}\kern-0.8em\TeX}}}

\setcopyright{acmcopyright}
\copyrightyear{2023}
\acmYear{2023}
\acmDOI{XXXXXXX.XXXXXXX}

\usepackage[utf8]{inputenc}
\usepackage{url}
\usepackage{lscape}
\usepackage{pifont}
\usepackage{dirtytalk}
\usepackage{makecell}
\usepackage{xcolor}

\usepackage{fancyvrb}
\usepackage{fvextra}

\usepackage{cleveref}[2012/02/15]
\crefformat{footnote}{#2\footnotemark[#1]#3}

\fvset{breaklines=true}
\RecustomVerbatimCommand{\VerbatimInput}{VerbatimInput}%
{fontsize=\footnotesize,
 frame=lines,  % top and bottom rule only
 framesep=2em, % separation between frame and text
 rulecolor=\color{gray},
 labelposition=topline,
}

\title{Targeted Attacks: Redefining Spear Phishing and Business Email Compromise}

\begin{document}

\author{Sarah Wassermann}
\affiliation{%
  \institution{Vade}
  \country{France}
}
\email{sarah.wassermann@vadesecure.com}

\author{Maxime Meyer}
\affiliation{%
  \institution{Vade}
  \country{France}
}
\email{maxime.meyer@vadesecure.com}

\author{S{\'e}bastien Goutal}
\affiliation{%
  \institution{Vade}
  \country{USA}
}
\email{sebastien.goutal@vadesecure.com}

\author{Damien Riquet}
\affiliation{%
  \institution{Vade}
  \country{France}
}
\email{damien.riquet@vadesecure.com}

\renewcommand{\shortauthors}{Wassermann et al.}

%%
%% The abstract is a short summary of the work to be presented in the article.
\begin{abstract}
In today's digital world, cybercrime is responsible for significant damage to organizations, including financial losses, operational disruptions, or intellectual property theft. Cyberattacks often start with an email, the major means of corporate communication. Some rare, severely damaging email threats -- known as spear phishing or Business Email Compromise -- have emerged. However, the literature disagrees on their definition, impeding security vendors and researchers from mitigating targeted attacks. Therefore, we introduce \textit{targeted attacks}. We describe targeted-attack-detection techniques as well as social-engineering methods used by fraudsters. Additionally, we present \textit{text-based attacks} -- with textual content as malicious payload -- and compare \textit{non-targeted} and \textit{targeted} variants.
%In today's digital world, cybercrime is responsible for significant damage to organizations. The consequences of cybercrime are not limited to financial losses, but also include operational disruption, reputational damage, or theft of intellectual property. Email being the most prominent form of communication and collaboration between organizations, many cyberattacks start with an email. Among these email-borne cyberattacks, some rare threats -- known as spear phishing or Business Email Compromise (BEC) -- have emerged, causing severe damage to organizations. However, as we point out in this paper, the literature does not agree on the definition of spear phishing and BEC. This makes it very difficult for security vendors and researchers to conceive methods to mitigate targeted threats. Therefore, in this paper, we introduce the notion of \textit{targeted attacks}, which can be precisely tailored to specify their scope. We thoroughly investigate the state of the art of targeted-attack-detection techniques and the social engineering methods used by the fraudsters to effectively carry out their attacks. As a final contribution, we present \textit{text-based attacks} -- email-borne cyberattacks where the malicious payload is the textual content -- and compare \textit{non-targeted} and \textit{targeted} variants.
\end{abstract}



\keywords{email, targeted attacks}


\maketitle

% Intro

\section{Introduction}
\label{sec:intro}
Today, we accomplish more and more tasks online, be it shopping, booking vacations, or managing finance. While the Internet has made our lives much easier, it has also created a plethora of possibilities for cybercrime, allowing criminals to reach targets instantly all over the world using simple tools such as electronic messages. Criminals attempt to make profit by obtaining access to worthy information, such as login credentials and critical personal data. Cybercrime also plays a prominent role in world crises, such as the invasion of Ukraine~\cite{google2023}.

Individuals and organizations face many cyberthreats on a daily basis -- fraudulent robocalls, suspicious instant messages and texts, or emails with a questionable syntax urging to click on a link or to open an attachment. Indeed, many threats use email as the attack vector, to steal the victim's login credentials through a counterfeit website, or to install malware on the victim's electronic device, to name a few. While most fraudulent emails are sent in bulk to reach as many victims as possible -- as it is the case for \emph{phishing} -- there are however types of attacks that are extremely rare and target specific individuals, attacks usually referred to as \textit{spear phishing} or \textit{Business Email Compromise} (\textit{BEC}). According to \citet{infosec2015}, the term \emph{phishing} appeared in the 90s, while spear phishing has been around only since the 2000s. The term \textit{spear phishing} appears in a report published by a working group on mass marketing fraud in 2006~\cite{group2006}, while Kaspersky states that they observed the first spear-phishing attacks in 2007 in the context of an espionage campaign~\cite{kaspersky2013}. Introduced more recently, the term BEC is considered as spear phishing in the enterprise setting by a set of publications such as~\cite{cidon2019, FBI2021, agazzi2020}. In this paper, we are focusing on both spear phishing and BEC.

Spear phishing and BEC are causing substantial financial damage to individuals and companies. According to the FBI, they lead to an exposed loss of more than \$43 billion between 2016 and 2021~\cite{FBI2022} for approximately 240,000 incidents reported. In other words, this means that each incident lead to an average loss of about \$180,000. To highlight the danger of these threats even further, \citet{cisco2011} reports that a spear-phishing campaign potentially yields ten times more money for the attackers compared to a mass-phishing campaign, while sending out 1,000 times fewer emails. Even large companies such as Facebook and Google have been victims of BEC attacks~\cite{cnbc2019}. Moreover, those threats are growing extremely fast. In 2017, the FBI reported that the identified exposed losses increased by 2,370\% in just two years, between January 2015 and December 2016~\cite{FBI2017}.

The published numbers underline that these new threats need to be taken seriously. Individuals as well as organizations have to be prepared, whether through the implementation of sophisticated detection technologies, or through security-awareness training programs. To address this situation, security vendors also have to put significant efforts into the development of efficient detection technologies. In particular, many legacy technologies are unable to cope with the challenge of detecting extremely rare and previously unseen malicious emails. A good start for this would be to provide a precise definition of spear phishing and BEC. This is first and foremost crucial for the academic world and researchers. The definition of a precise research topic is essential, whether it is to identify the different relevant related publications, or to compare research results. It is also necessary for security vendors and their customers, in particular to clearly identify the capabilities of a product or a technology to detect certain types of threats. Unfortunately, the literature does not agree on those definitions, as we unveil in this paper. Therefore, our main goal is to provide the cybersecurity community with clear and detailed definitions of these threats.

The remainder of our paper is organized as follows. In Section~\ref{sec:contradictions}, we first present the widely accepted definition of \textit{phishing}, before we demonstrate that the notions of \textit{spear phishing} and \textit{BEC} are not well-defined in the literature. In Section~\ref{sec:targeted_attacks}, we introduce the notion of \textit{targeted attacks}, describe their characteristics and related detection techniques proposed in the literature. In Section~\ref{sec:text_attacks}, we focus on text-based attacks and present both non-targeted and targeted ones. Section~\ref{sec:conclusion} concludes this paper.
\section{Definitions and Contradictions across the Literature}
\label{sec:contradictions}

The definition of \textit{phishing} ($P$) is consistent across the literature. The following definition is provided by the National Institute of Standards and Technology (NIST), and many sources agree on it: \say{A technique for attempting to acquire sensitive data, such as bank-account numbers, through a fraudulent solicitation in an email or on a website, in which the perpetrator masquerades as a legitimate business or reputable person}~\cite{nist2022}. The term \textit{phishing} comes from \emph{fishing}, and alludes to the fact that \say{fishers (i.e., attackers) use a bait (i.e., socially-engineered messages)
to fish (e.g., steal personal information of victims)}~\cite{khonji2013}. Phishing attacks share a number of common features:

\begin{itemize}
    \item The attack relies on the impersonation of a trustworthy entity: usually an organization, and, more rarely, an individual. In particular, phishing usually impersonates commercial brands -- such as financial services (PayPal, Chase, etc.), cloud services (Microsoft, Google, etc.), logistics/shipping (FedEx, DHL, etc.), online retail (Amazon, Alibaba, etc.) and social media (LinkedIn, Facebook, etc.) -- or government agencies (such as the Internal Revenue Service)~\cite{apwg2022, vade2022}.
    \item The attack starts by sending a message, usually an email, or sometimes an instant message (WhatsApp), a text message (\textit{smishing}), or even by making a phone call (\textit{vishing}). The message contains a text that leverages social engineering techniques (e.g., creating a sense of urgency) to urge the user to click on a link that redirects them to a fraudulent website. Both the message and the website contain textual and visual clues (logo, visual identity, look and feel) to impersonate the trusted entity.
    \item The goal of the attacker is to acquire sensitive data such as login credentials, bank-account numbers, and critical personal information (such as the Social Security Number).
\end{itemize}

Furthermore, phishing messages are sent in bulk to large lists of recipients, to reach the maximum number of people before the attack is detected and blocked, and this has several consequences. First, since a phishing message is not personalized, it can be easy for the recipient to detect its fraudulent nature. For example, an individual may receive a message pretending to originate from a particular bank (e.g., Wells Fargo) -- claiming a payment issue or a bank-account overdraft -- while the individual is not a customer of this bank. In this case, the fraud is easily detected as it appears out of context. Therefore, a phishing attack rarely succeeds, and if it succeeds, the impact is usually limited from a financial point of view. Indeed, regarding phishing attacks, the FBI reported for 2021 an accumulated loss of \$44,213,707 for 323,972 victims: the average loss per victim is about \$136.47~\cite{FBI2021}. Additionally, as phishing is a common cyberthreat, it is quite easy to collect samples (phishing messages, fraudulent websites, phishing kits) to analyze them, and thus design and develop technologies to detect these attacks. From an academic point of view, there is indeed abundant literature on this topic, and many detection methods are well documented (blacklists, heuristics, visual similarity, data mining, machine learning)~\cite{khonji2013, almomani2013, sahoo2017}.

Whereas phishing campaigns -- similar to spam and malware campaigns -- are sent indiscriminately to large lists of recipients, new types of low-volume email-borne cyberthreats have emerged, with the particularity of targeting specific individuals, and with a high level of personalization to increase the probability of deceiving them. New terms have been coined to describe these targeted fraudulent emails. The first term is \textit{spear phishing} (\textit{SP}), derived from \textit{phishing} and \textit{spearfishing}, that emphasizes the fact that it is a targeted \textit{phishing} attack, using the metaphor of a spear. Initially, spear phishing has been used in particular to describe email-borne cyberattacks that aim at infiltrating an organization by targeting one or several of its employees. The second term, that appeared later, is \textit{Business Email Compromise} (\textit{BEC}), where the attacker typically tricks a business into carrying out a wire transfer to a fraudulent banking account, claiming for example an outstanding invoice from a supplier \citet{FBI2017}. The difference between the terms spear phishing and BEC is unclear, confusing, and evolves over time both in the cybersecurity industry and in the literature.

% Introduce table
% The table is key to demonstrate that the definitions of SP and BEC are inconsistent in the literature. Each feature will then be reviewed one by one to support the demonstration.
Table~\ref{tab:paper_comparison} presents the terminology used in the literature, as well as the characteristics associated to the definition of each term. Publications are sorted by date, from the oldest to the most recent. Column `\textbf{Term}' contains the term used in a publication (\textit{BEC}, \textit{SP}, \textit{TA} for \say{Targeted Attack}, and \textit{BEC P} for \say{Business Email Compromise Phishing}). Column `\textbf{Attack Vectors}' describes the attack vectors used to carry out the attack, whether it is a URL (leading to a fraudulent website), an attachment (containing a malware), or the textual content of the email (to trick the victim into performing an action through social engineering techniques). Column `\textbf{Attack goals}' lists the different attack goals considered in each publication (refer to Table~\ref{tab:attack_goals} presented in Section~\ref{sec:targeted_attacks} for a description of these goals). Column `\textbf{Targets}' identifies the typical targets of the attack. Finally, Column `\textbf{Who is impersonated?}' gives more details about the different impersonation schemes. It is important to note that the definition of the term used (SP, BEC) is rather vague in most publications, and that some identified characteristics were inferred from the context of the paper.

% Chronological point of view: SP has been somehow superseded by BEC
The first observation is that -- from a chronological point of view -- the term \textit{BEC} appeared later than the term \textit{SP}, and that the term \textit{SP} is rarely used nowadays. There are two hypotheses. The first one is that these two terms refer to two distinct email-borne cyberthreats, and the term \textit{SP} is less used because the associated threat is less prominent than before. The other hypothesis is that these two terms refer to the same email-borne cyberthreat, and that \textit{BEC} is becoming the customary name for this kind of fraud.  Based on our experience in the  cybersecurity industry, we can suppose that the fact that the term \textit{BEC} is used by the FBI -- while the term \textit{SP} is not -- has led the industry and academics to adopt this designation~\cite{FBI2017,FBI2021,FBI2022}. Note that the term \textit{targeted attack} appeared early in the literature, as a first definition was proposed by Cisco in 2011~\cite{cisco2011}. The term, however, has not persisted in the state of the art. \citet{cisco2011} defines a \textit{targeted attack} as a \say{highly customized threat}, typically based on a malware, that aims to infiltrate an organization or steal intellectual property. This definition differs from ours which is broader. In particular, the definition of \citet{cisco2011} fails to capture most targeted attacks observed today.

% Attack Vectors: We can find a trend (more text-based attacks), but there is an obvious lack of consistency
Regarding the attack vectors, Table~\ref{tab:paper_comparison} shows that there is a growing number of text-based attacks over time. Indeed, we can see from the table that, while URLs and attachments are the main attack vectors of spear phishing, BEC highly relies on a carefully crafted textual content, leveraging social engineering techniques. \citet{abnormal2022} explains this evolution of the attack vector over time: \say{What all these email attacks have in common is that they are almost exclusively text-based in an effort to bypass traditional email security tools that look for malicious attachments and links.}. There are however inconsistencies across the literature, both for SP and BEC. \citet{cisco2011} mentions that 80\% of SP attacks are URL-based, while \citet{trendMicro2012} reports that 94\% of SP attacks are attachment-based. Regarding BEC, some publications consider only text-based attacks~\cite{agazzi2020,agari2020,agari2021}, while others include URL-based and attachment-based attacks~\cite{blond2014, trendMicro2018,cidon2019,greathorn2021,almusib2021}. Numbers also vary in the literature. While \citet{cidon2019} report that 40.1\% of BEC attacks include a malicious link, \citet{almusib2021} claim that a link or an attachment is used in only 3\% of the cases.

% Attack goals
% SP : Mostly GC3, sometime GC2
% BEC: Mostly GC1
% Once again, there are inconsistencices
As for the purpose of the attack, the literature once again presents variations, as illustrated in Table~\ref{tab:paper_comparison}. Publications on SP focus on the infiltration of organizations (GC3) through a malicious link or attachment, sometimes combined with the theft of confidential documents and critical information (GC2). A notable exception is \citet{cisco2011}, where the goal of SP is only financial (GC1). Regarding BEC, the goal is primarily financial (GC1), via fraudulent wire transfers. In this respect, the term BEC can be misleading because it first evokes the compromise of a business (GC3), as defined in~\cite{FBI2017, FBI2021}. However, \citet{cidon2019} report that the goal of 46.9\% of BEC attacks is a wire transfer (GC1), 12.2\% to establish a rapport with the target (GC4), and 0.8\% to steal PII (GC2).

% Targets are organizations, and the individuals in these organizationsliterature
% No major inconsistency here
The literature is consistent with the fact that SP attacks target individuals and organizations, where the organization is typically a business (Google, RSA~\cite{ho2017}), but may also be a government agency (US State Department, White House~\cite{ho2017}), a political party (DNC~\cite{ho2017}) or a non governmental organization (WUC~\cite{blond2014}). In this context, individuals refer to members of these organizations, and do not refer to the general public, as it is very unlikely that individuals in the general public have access to very valuable assets and thus be targeted by SP attacks. It is important to note that individuals include high-profile political figures, such as John Podesta and Hillary Clinton~\cite{ho2017,ilangakoon2018}. In some cases, the individuals targeted are senior executives and high-rank staff, a specific type of SP attack known as \textit{whaling} ('phishing' the big `fishes')~\cite{butavicius2016, salahdine2019}. In the case of BEC, the primary target are businesses, and certain roles within the organization are more targeted than others. \citet{cidon2019} report that C-level executives (including CEO and CFO) are targeted in 29.3\% of the cases, finance/HR staff in 16.9\% and other employees in 53.7\%. \citet{agari2020} reveals that, in the case of a BEC campaign named 
'Cosmic Lynx', almost all employees targeted were senior-level executives (CEO, CFO, President, Vice President, General Manager, Managing Director).

% Impersonation
The literature shows that SP attacks rely quite often on the impersonation of a trusted sender. The trusted sender can be, for example, a work colleague or a business partner~\cite{stringhini2015}. Regarding BEC attacks, there is a large consensus in the literature that impersonation -- usually leveraging a spoofing technique and more rarely based on a compromised email account -- plays a key role to carry out text-based targeted attacks that rely mostly on social engineering techniques. The impersonated individual may be someone belonging to the organization (an executive or an employee) or someone external to the organization (a supplier, a customer, or a lawyer, identified with an asterisk in the column `\textbf{Who is impersonated?}'). \citet{cidon2019} provide some statistics in this regard. Inside the organization, 49.6\% of impersonated individuals are C-level executives, while 50.4\% are other employees.

% Comparison of SP and BEC in literature, a major contribution to the demonstration as the comparison of SP vs BEC is not consistent.
While most publications refer to either SP or BEC, only a handful refer to both and compare one to the other. We note a significant lack of consistency between the different presented definitions. \citet{salahdine2019} define spear phishing attacks and \say{business-email-compromise-phishing} attacks as two different categories of phishing attacks. The goal of the \textit{BEC-phishing} attack is to get access to the target's business email account. \citet{cidon2019} present BEC and whaling (spear phishing targeting high-rank staff) as equivalent. \citet{greathorn2021} proposes a different definition, and spear phishing is considered as a particular type of BEC attack. Oppositely, \citet{almusib2021} refer to BEC as being a specific type of spear phishing. The definition of spear phishing by the NIST further underlines this problem by being very vague: \say{A colloquial term that can be used to describe any highly targeted phishing attack}~\cite{nist2022b}.

\newcolumntype{C}[1]{>{\centering\let\newline\\\arraybackslash\hspace{0pt}}m{#1}}

%\begin{landscape}
%\topskip0pt
%\vspace*{\fill}
\begin{table}
    \centering
    \small
    \begin{tabular}{|c||c|C{2.2cm}|C{1.8cm}|c|C{3cm}|}
    \hline
        \textbf{\makecell{Literature}} & \textbf{\makecell{Term
}} & \textbf{\makecell{Attack vectors}} & \centering \textbf{Attack goals} & \textbf{Targets} & \textbf{Who is impersonated?} \\ \hline \hline
        \citet{cisco2011} & SP & URL (80\%) & GC1 & Org. & ? \\
        ~ & TA & Attach. & GC2, GC3 & Org. & ?\\ \hline
        \citet{trendMicro2012} & SP & Attach. (94\%) & GC3 & Org. & ? \\ \hline
        \citet{parmar2012} & SP & URL, attach. & GC3 & Org. & Trusted source\\ \hline
        \citet{blond2014} & TA & Attach. & GC3 & Org. & High-profile identities known by the victims\\ \hline
        \citet{caputo2014} & SP & URL & GC3 & Org. & Well-known companies, trusted relationships\\ \hline
        \citet{stringhini2015} & SP
 & URL, attach. & GC2, GC3 & Org. & Work colleague, business partner\text{*}\\ \hline
        \citet{halevi2015} & SP & URL & GC3 & Org. & Trustworthy source\\ \hline
        \citet{han2016} & SP & Attach. & GC2, GC3 & Org. & ?\\ \hline
        \citet{duman2016} & SP & URL, attach. & GC3 & Org. & Trusted senders \\ \hline
        \citet{zhao2016} & SP & URL, attach. & GC2, GC3 & Org. & ?\\ \hline
        \citet{butavicius2016} & SP & URL
 & GC2, GC3 & Org. & Legitimate entity\\ \hline
         \citet{FBI2017} & BEC & Text & GC1, GC2 & Org. & Executive, employee, supplier\text{*}, attorney\text{*}\\ \hline
        \citet{ho2017} & SP & URL & GC3 & Org. & Trusted entity  \\ \hline
        \citet{gascon2018} & SP & URL, attach.
 & GC3 & Org. & ? \\ \hline
        \citet{ilangakoon2018} & SP & URL, attach. & GC2, GC3 & People & ?\\ \hline
        \citet{trendMicro2018} & BEC & URL, attach., text & GC1, GC2, GC3 & Org. & Executive, lawyer\text{*}\\ \hline
        \citet{burns2019} & SP & URL, attach., text & GC1, GC2, GC3 & Org. & Legitimate actors\\ \hline
        \citet{salahdine2019} & SP & URL, attach. & GC1, GC2, GC3 & Org. & ?\\ 
        ~ & BEC P & URL, attach. & GC1, GC2, GC3 & Org. & ?\\ \hline
        \citet{cidon2019} & BEC & URL (40.1\%), attach., text& GC1, GC2, GC3, GC4& Org. &  CEO (42.9\%), CFO (2.2\%), C-level (4.5\%), finance/HR (2.2\%), employees (48.1\%)\\ \hline
        \citet{ho2019} & SP & ? & ? & Org. & Known and legitimate user\\ \hline
        \citet{agazzi2020} & BEC & Text & GC1, GC2 & Org. & Executive, lawyer\text{*}\\ \hline
        \citet{agari2020} & BEC & Text & GC1 & Org. & Executive, lawyer\text{*}\\ \hline      \citet{regina2021text} & BEC / SP & Text & GC1, GC2, GC4 & Org. & Executive, employee\text{*}\\ \hline       
        \citet{agari2021} & BEC & Text & GC1 & Org. & Executive, employee\\ \hline        
        \citet{greathorn2021} & BEC & URL, attach., text & GC1, GC2, GC3 & Org. &  Manager, employee, customer\text{*}, company\text{*}\\ \hline
        \citet{almusib2021} & BEC & Text (97\%) & GC1, GC2 & Org. & Executive, employee, supplier\text{*}, lawyer\text{*}\\ \hline
        \citet{FBI2021} & BEC & Text & GC1, GC2 & Org. & Executive, vendor\text{*}\\ \hline
        \citet{evans2022} & SP & URL, attach., text & GC1, GC3 & Org. & Trusted individual\\ \hline
    \end{tabular}
    \caption{Summary of the attack characteristics found in the literature. An asterisk identifies an impersonated indivdual external to the organization. More details about the goals are presented in Table~\ref{tab:attack_goals}.}
    \label{tab:paper_comparison}
\end{table}
%\vspace*{\fill}
%\end{landscape}

% Introduction of the notion of targeted attack
\section{Targeted Attacks}
\label{sec:targeted_attacks}

% Goal of this paragraph: Propose a term that can be refined with adjectives. It is self-explanatory and consistent.
To remedy the problem of definition, we introduce in this paper the term \textit{targeted attack}, to replace the vague notions of \textit{spear phishing} and \textit{BEC}. We consider that this definition is different from the one proposed in 2011 by Cisco~\cite{cisco2011} as it is not referring to the same threat. The term \textit{targeted attack} is opposed to the one of \textit{non-targeted attack}. We can fine-tune these two notions (targeted and non-targeted) by adding adjectives to reduce the scope. 
We can list the three following examples for clarity purposes: 
\begin{itemize}
    \item \textit{Attachment-based attacks}: these are attacks with an attachment containing a malicious payload -- such as a malware, or a downloader that fetches and installs a malware. These attacks may be targeted (\textit{attachment-based targeted attacks}) or not (\textit{attachment-based non-targeted attacks}) ~\cite{trendMicro2012}.
    \item \textit{Corporate targeted attacks}: they correspond to targeted attacks specifically aiming corporations. They include attacks where the malicious payload is the textual content (\textit{text-based corporate targeted attacks}, such as payroll fraud or CEO fraud) or attacks where the malicious payload is included in an attachment (\textit{attachment-based corporate targeted attacks}).
    \item \textit{URL-based non-targeted attacks}: they refer to non-targeted attacks where the attack vector is a URL included in the message, which points to a malicious file or a fraudulent website (e.g., a phishing attack)~\cite{trendMicro2012}.
\end{itemize}
We detail the notion of \textit{targeted attack} in the rest of this section and provide an overview of different targeted-attack detection techniques presented in the literature.

% First, who is targeted by targeted attacks?
\paragraph{\textbf{The attack targets specific individuals or employees.}} A targeted attack is a fraudulent electronic message (email, text message, instant message) that targets either a specific individual, or an organization (a corporation, an enterprise, a business, a government agency, an educational institution, etc.) through one or several specific individuals that have a key role within the targeted organization. \citet{greathorn2021} reported in 2021 that employees working in finance departments were targeted in 57\% of the cases, followed by the CEO (22\%) and the IT department employees (20\%). The choice of the individuals targeted is motivated by the access to sensitive resources (e.g., bank accounts, information system, contracts) and the position of authority within the organization (e.g., C-suite executive) in adequacy with the goal of the attack defined in Table~\ref{tab:attack_goals}.

% Second, who is impersonated?
\paragraph{\textbf{The attacker impersonates a trusted party.}} A targeted attack typically relies on the impersonation of a trusted party of the victim (see column~\textbf{'Who is impersonated?'} in Table~\ref{tab:paper_comparison}). It may be a family member, a relative, or a friend of the victim. In the context of a corporate attack, it might be a C-suite executive (CEO, CFO), a contractor, a supplier, a customer, or a partner~\cite{FBI2021, trendMicro2018}. The impersonation can be carried out by compromising directly the means of communication of the impersonated individual. For example, compromising the email account is ideal for the attacker, as existing email conversation threads can be hijacked to carry out targeted attacks -- attacks that are very challenging to detect as the email address is authentic. However, in practice, it is very difficult for the attacker to compromise an email account. For this reason, impersonation is usually performed by sending a message containing visual cues to trick the victim, e.g., first name and last name of the impersonated individual in the sender field of the message~\cite{cidon2019, greathorn2021}. In some cases, the attackers imitate the writing style of the impersonated individual to make the attack even more credible~\cite{gascon2018}. In contrast, non-targeted attacks (phishing, spam) appear most of the time out of context to the victim, as there is no relation between the sender and the recipient.

% Third, it is a tailored attack
\paragraph{\textbf{Targeted attacks are tailored to the victim.}} An important characteristic is the tailored nature of the attack. Indeed, in the context of a corporate targeted attack, the attacker not only knows the first name and last name of the victim, but also knows their role within the organization. The most sophisticated targeted attacks contain information that shows an in-depth knowledge of the organization (knowledge of the organizational chart, financial processes within the organization, suppliers and customers, etc.). 

% Attack vector
\paragraph{\textbf{The attack can be executed through different vectors.}} The \textit{attack vector} (or malicious payload) of the targeted attack can take different forms. Some targeted attacks rely on textual content only, urging the victim to perform an action (wire money, disclose confidential information) by leveraging social engineering techniques. This is more and more the case, as demonstrated by Table~\ref{tab:paper_comparison}. We define these attacks as \textit{text-based targeted attacks}. The attack vector can also take the form of a URL (to lead the victim to a fraudulent website) or the form of an attachment (e.g., malware, ransomware).

% Goal & types of the targeted attacks
\paragraph{\textbf{The attacker wants to fulfill a specific goal.}} A targeted attack is also characterized by its goal, e.g., steal money, obtain confidential information, or infiltrate an organization (see Table \ref{tab:attack_goals}). Some techniques are associated to a specific goal. For instance, the attacker can get money from the victim through different means, such as invoice fraud, payroll fraud, or ransomware. The attack technique may be associated with the attack vector. For example, an invoice fraud is often performed with an attached PDF invoice that contains the fraudulent bank-account details, while credential theft relies on a URL sent in an electronic message (email, text message, instant message) that redirects the victim to a fraudulent website that mimics the login page of a trusted online service. A targeted attack may also be the first step of a multi-stage attack. A first innocuous email -- that may appear non-targeted at first glance -- may be sent to the victim, to establish a trust channel, and to also potentially disarm certain email-security defense layers that rely on historical communications. The malicious payload is sent only after several emails have been exchanged. Similarly, a single-stage or multi-stage targeted attack can also be the first step of a larger attack. For example, the attackers may steal information to access the network of a company to perform a Denial of Service (DoS) attack later on~\cite{stringhini2015, gascon2018, trendMicro2012}.

% Targeted: Low volume of attacks
\paragraph{\textbf{Targeted attacks are extremely rare.}} Another characteristic of targeted attacks is their rarity. As already stated in Section~\ref{sec:intro}, it is crucial to emphasize that targeted attacks occur extremely rarely compared to other, less targeted attacks. This observation is strengthened by volumetric numbers reported by security vendors. \citet{abnormal2022} reveals that about 76.7\% of attacks targeting organizations were phishing attacks in the second semester of 2021, while only 3.81\% were targeted attacks. In the same vein, they found an average of 0.82 targeted attacks per 1,000 mailboxes during the same period. For the sake of comparison, Valimail observed in 2019 that more than 3 billion fraudulent emails are sent every day~\cite{valimail2019}. The rarity of targeted attacks is also highlighted in multiple research papers. Indeed, \citet{cidon2019} report that they spotted less than 1 out of 50,000 emails to be a targeted attack. \citet{stringhini2015} rely on a dataset composed of more than 187,000 emails; 550 of them are considered as targeted corporate attacks (1 out of 340 emails). The corporate dataset presented by \citet{ho2017} is an even more striking example: while containing more than 370 million emails, it encompasses merely at most 10 targeted attacks. \citet{han2016} use a dataset incorporating a much larger proportion of malicious samples compared to other publications, with 1,467 targeted attacks to 14,043 benign samples. This is however a deliberate choice of \citet{han2016} in order to reduce the identified class-imbalance issue (See Section~\ref{subsec:targeted_attacks_detection}).

\begin{table}[!ht]
    \centering
    \hspace{-1cm}
    \begin{tabular}{|c||C{10cm}|}
    \hline
        \textbf{Goal} & \textbf{Description} \\
        \textbf{category} & \\        
        \hline \hline
        \textbf{GC1}& The goal is financial, i.e., obtain money from the victim. This can be perpetrated via different text-based targeted attacks, such as CEO fraud, gift card scam, or payroll fraud (Section~\ref{sec:text_attacks}). \\ \hline
        \textbf{GC2} & The goal is to obtain confidential or critical information from the victim, such as login credentials, sensitive Personal Identifiable Information (PII) (e.g., Social Security Number), contracts, or intellectual property assets (e.g., formulas, practices, processes). Credential theft is typically achieved via URL-based attacks. W-2 fraud is an example of a text-based targeted attack that aims at retrieving Social Security Numbers of the organization's employees, in order to perpetrate other types of fraud, such as identity theft (Section~\ref{sec:text_attacks}).  \\ \hline
        \textbf{GC3} & The goal is to infiltrate the infrastructure of the organization. This can be achieved by compromising an email account within the organization. It may be the first stage of a sophisticated covert attack, that includes reconnaissance, espionage, and exfiltration of high-value data. \\ \hline
        \textbf{GC4} & The goal is to establish a trust channel with the recipient, and possibly disarm security technologies based on historical communications. This can be achieved by sending one or several innocuous messages to the recipient. This is typically the first stage of a multi-stage attack. \\ \hline
    \end{tabular}
    \caption{The different goals of targeted attacks.}
    \label{tab:attack_goals}
\end{table}

% Social engineering techniques
\paragraph{\textbf{Attacks leverage social engineering techniques.}}
To perform targeted attacks, the attackers make use of sophisticated social engineering techniques~\cite{agazzi2020, butavicius2016, gonzalez2015, jagatic2007, allodi2020}. \citet{agazzi2020} and \citet{uebelacker2014} present the Big-Five framework traits, which are the five personality traits that might characterize the victim and increase the likelihood of being tricked by a social engineering attack, namely:
\begin{itemize}
    \item \textit{Conscientiousness}: the character trait of wanting to carry out every task well and of taking every order seriously.
    \item \textit{Extraversion}: the personality trait of being outgoing and enjoying social interactions.
    \item \textit{Agreeableness}: an agreeable person puts the needs of others before their own.
    \item \textit{Openness to experience}: a person with this character trait enjoys new experiences, is creative, and is intellectually curious.
    \item \textit{Neuroticism}: a person suffering from high neuroticism is more prone to depression.
\end{itemize}

Besides these personality traits, the attackers may use some principles to manipulate and influence their victims. \citet{agazzi2020, cialdini2009, sharevski2022, taib2019, debona2020}, and \citet{uebelacker2014}, describe six of them, namely:
\begin{itemize}
    \item \textit{Authority}: people tend to comply to orders and decisions coming from people seen as holding power or authority over them, even if they disagree with them, as they might fear the consequences of disobedience. Authority in emails can be perceived through different ways: the sender might indicate their position in the company by mentioning it in the signature, or use a writing style that does not allow any room for negotiation (formal and imperative tone). People being experts in their domain -- such as a lawyer, a judge, or a policeman for the legal domain, a pastor, or a priest for religion -- can also be considered as people exerting authority. This principle is often used in the context of a lawyer and CEO frauds. When the CEO requests a certain task to be done quickly, for instance, the employee will most probably execute it without asking many questions.
    \item \textit{Social proof}: people are more likely to perform a certain action if it is \say{socially accepted}, in other words, if many other people in their surroundings have carried out the same action. This circle of people can be different according to the context: family, friends, work colleagues, or even strangers physically around them in a close environment (waiting room, restaurant, stadium). For example, people are more likely to buy a product which has been bought by thousands of other individuals. In general, people trust more fellow humans with similar opinions, especially in ambiguous situations. Fraudsters rely on this principle to present to the victim an unfamiliar situation, for which they have no prior reference or training. In the case of targeted attacks, in the absence of a reference other than the message, the victim will rely on the latter to determine how to act.
    \item \textit{Scarcity}: people tend to value more scarce resources or rare opportunities. For instance, people will hesitate less before buying a product which is advertized as very limited by a popular shop. In general, this principle appeals to the fear of missing out on something. It also applies to requests which claim to be urgent, for which one must act quickly, even when the deadline is implicit. The emergency markers influence people to comply with a request without taking the time to reflect on it. In the context of frauds, the attacker applies this principle when impersonating the IT department and asking for an urgent password renewal, before the end of the working day, for example, or targets a member of the sales team hinting that a potential deal could be made quickly.
    \item \textit{Commitment and consistency}: people attach importance to their actions being consistent with the values they believe in. An individual who has lost a loved one because of cancer is more likely to donate to charity for this illness, for example. Moreover, once people have taken a public stand, they feel pressured to stick to it. Similarly, this principle applies to habits and routine actions. This is leveraged in messaging attacks by attackers being consistent in their demand, often starting with small requests and then asking for more. This pattern can also be seen in some targeted attacks. Indeed, once the victim responded to the first solicitation, they are more inclined to respond to the following ones.
    \item \textit{Liking}: people tend to be convinced more easily by people they know, like, or are similar to (professional situation, sickness, family status, hobbies...). The same goes for actions: people are more likely to perform an action they enjoy doing. For instance, teenagers often want to adopt the same lifestyle as their idols. In case of an attack, the fraudsters exploit this manipulation principle and make their message or request more likeable. Phishing attempts will be more convincing if the message is visually appealing and usurps a popular brand. In the case of targeted attacks, they might impersonate a close friend or a work colleague of the victim.
    \item \textit{Reciprocity}: people often respond to a positive action with another positive action. Indeed, reciprocity is based on the universal understanding that people give back to the ones who have given first. For instance, a person is more likely to do a favor for a friend who has been helpful in the past. It is important to note that the initial gesture does not need to be a physical gift: it can be a smile, a compliment, or information. However, the gesture must be valuable to the recipient. In the context of a fraud, the attacker might promise to give access to confidential information if a certain attachment is downloaded or a link is clicked on.
\end{itemize}

\citet{agazzi2020} demonstrates how these influential principles were applied in the targeted attacks against Ubiquity and Peebles Media Group. Previous research shows that people have a lot of difficulties detecting targeted attacks when the email contains elements of the authority principle~\cite{butavicius2016}. The use of these techniques in text-based attacks is further highlighted with different annotated examples in Section~\ref{sec:text_attacks}.

Some authors emphasize the fact that social engineering techniques make use of human emotions. In the context of social engineering techniques used to propagate malware, \citet{abraham2010} list the following emotions: \textit{curiosity}, \textit{empathy}, \textit{excitement}, \textit{fear}, and \textit{greed}.
 Fear is often leveraged, based on the fact that the Internet is perceived as unsafe by the end users. A particular type of malware known as \textit{scareware} exploits the end users' fear and anxiety by displaying fake pop-up warnings (computer infected, fraud alert, disk corruption, etc.) to convey them to click on a link to download the offered software solution. They may end up purchasing an unnecessary software. In the worst-case scenario, a malware may be installed on the computer. Regarding greed, it is common that end users are tricked into opening an email or visiting a website under the pretense of an opportunity. Opportunities can take different forms -- employment, investment, lottery, etc.

\paragraph{\textbf{The attacker gets to know their victim.}}
To find the necessary information about the target (be it an individual or an organization), the attackers often carry out an in-depth analysis of the target's online presence~\cite{salahdine2019, jagatic2007, almusib2021}. The website of an organization typically lists the senior leadership positions within the organization, and provides the attacker with a list of potential high-ranking individuals to impersonate or target. Social-networking platforms are also leveraged by the attacker to retrieve additional information~\cite{parmar2012, almusib2021, jagatic2007, burns2019}. Professional social networks such as LinkedIn can provide a list of employees of an organization with employment-oriented information -- the position within the organization, experience, education and professional network. Other social networks (e.g., Facebook) provide more personal details about these employees (family and relatives, friends). Information published online by the organization and its employees through different channels (website, blogs, social networks, online media) can also be used to build a realistic context for targeted attacks. For instance, the participation of an organization to a trade show that is advertised on social media may lead to a fraudulent-invoice targeted attack asking the payment of certain services related to this trade show. In the case where the attacker has gained access to an email account inside the organization, they can extract highly relevant information (e.g., writing style of the employee, examples of invoices, upcoming events) from different sources (e.g., emails, calendar, documents) to craft even more convincing targeted attacks launched from the inside. The technique of launching attacks from the inside is also referred to as a \textit{lateral attack}~\cite{ho2019}. Some of these attacks may also use an existing conversation thread, a technique known as \textit{conversation hijacking}~\cite{barracuda}.

\paragraph{\textbf{The attacker develops a method to execute their attack plan.}}
Once the attacker has gathered enough information to craft a convincing targeted attack, the next step is to send the message to the victim posing as a trusted individual. In the case where the means of communication of the impersonated individual has been compromised (email-account compromise), it is straightforward for the attacker to send the message. However, email-account compromise is marginal, and accounts only for 4\% of targeted attacks as reported by \citet{greathorn2021}. In most cases, the attack is launched from the outside of the organization, and the attacker has to leverage techniques specific to the type of electronic message (email, text message, instant message) to \textit{spoof} the identity of the sender. In the case of email, \textit{spoofing} is achieved by forging the \texttt{From} header, as this header contains the display name and email address of the sender~\cite{opazo2017, blond2014, greathorn2021}. According to a report published by \citet{greathorn2021}, the attackers spoofed an identity in the display name in 49\% of the attacks in 2021. This report also revealed that 18\% of targeted attacks rely on look-alike domains to impersonate the sender email address, a technique also observed by other authors~\cite{cidon2019, blond2014}. Free webmail accounts are often abused to carry out targeted attacks, as they do not control the identity of the user and are usually trusted by the recipient. \citet{agari2021} found that around 77\% of the targeted attacks originated from free webmail accounts in the second semester of 2020, with 61\% from Gmail only.

\subsection{Detection of Targeted Attacks}
\label{subsec:targeted_attacks_detection}

The literature encompasses numerous papers presenting different approaches to detect email-borne targeted attacks. However, the scope of the results depends on the quantity and quality of the data available for the experiments. There are only a handful of public email datasets available, and these datasets are small and sometimes outdated (as the Enron corpus~\cite{klimt2004}, for example). The rarity of targeted attacks has to be considered as it results in an extreme class-imbalance problem. Therefore, it is necessary to run experiments on very large imbalanced datasets to estimate the false-positive rate~\cite{ho2017, cidon2019}. \\

\citet{stringhini2015} and \citet{duman2016} propose to build a user profile to enhance fraud detection. \citet{stringhini2015} present IdentityMailer, a system that analyzes and filters outgoing emails on the sender-side. To do so, IdentityMailer builds a behavioral profile of the sender based on three categories of features: \textit{writing habits}, \textit{composition and sending habits}, and \textit{interaction habits}. In the case an email is flagged as suspicious, the sender needs to answer a questionnaire to prove their identity. The system is however impractical as the false-positive rate is in the 1-10\% range. \citet{duman2016} introduce EmailProfiler, which aims at determining if an email is actually from the sender claimed in the email metadata. To this end, EmailProfiler builds a sender behavioral profile based on features extracted from the textual content (lexical, syntactic and structural features) and from headers of past emails received by the user. An incoming email can then be compared to the profile of the sender to determine whether it is genuine or not. Alternatively, the user can also generate their own email-sending behavioral profile which can then be shared with others via a trusted server. 

\citet{han2016} present a semi-supervised learning model coupled with an affinity graph to detect spear phishing and, as a second step, to attribute the attack to a known spear-phishing campaign. To this purpose, four categories of features are extracted from the email: \textit{origin features} (e.g., source IP address and Autonomous System Number (ASN)), \textit{text features} (e.g., topics), \textit{attachment features} (e.g., file type), and \textit{recipient features}. Two datasets are used to evaluate the technique: a spear phishing dataset of 1,467 emails, and a benign dataset of 14,043 emails. While class imbalance is taken into account, it is however much less pronounced than in the wild~\cite{ho2017}.

\citet{zhao2016} discuss the problem of tuning the threshold to classify an email so that it is personalized for each individual. This is especially important in the context of an organization where users have significant differences in their susceptibility to spear-phishing attacks as well as access to critical credentials. Different scenarios are studied (\textit{single-credential} and \textit{multiple-credential} scenarios) and optimal defense strategies formulated as optimization problems are presented. 

\citet{ho2017} present a real-time credential-stealing spear-phishing detector. The detector leverages historical data, such as past emails, user-login information, and enterprise network-traffic logs. To evaluate their detector, Ho et al. rely on a dataset encompassing 370 million emails, with less than 10 known spear-phishing attacks. As standard machine-learning techniques are likely to fail due to the extreme class imbalance, they implemented a novel anomaly-detection technique called Directed Anomaly Scoring (DAS) with the goal of achieving a false-positive rate that is low enough to be practical. Two sets of features are extracted from the email: \textit{domain-reputation features} to determine the risk associated to a link, and \textit{sender-reputation features} to detect different types of impersonation. Their work is however restricted to URL-based targeted attacks.

\citet{gascon2018} designed an anti-spoofing framework based on learned content-agnostic sender profiles, i.e., the textual content of the email is ignored. Three types of features are considered to build a sender profile: \textit{behavior features} capturing the sender's preferences and peculiarities (types of attachments, use of the \texttt{Cc} header, etc.), \textit{composition features} that identify a particular email-client software along with its configuration parameters (encoding schemes, structure of MIME parts and boundaries, etc.), and \textit{transport features} that reflect the delivery path of the email from a technical point of view (Autonomous System Number of source IP, results of DKIM validation, TLS features, etc.). 

\citet{cidon2019} present BEC-Guard, a system that can detect BEC with a very low false-positive rate. To address the class-imbalance problem, the classification pipeline is composed of chained classifiers. The first classifier aims at detecting impersonation, based on email headers and historical communication patterns. If an impersonation is detected, then a set of two content classifiers is applied. The first content classifier (\textit{text classifier}) processes the textual content and produces a feature vector with the term frequency-inverse document frequency (TF-IDF) algorithm. The second content classifier (\textit{link classifier}) extracts features from the links. The impersonation and link classifiers use random forests, while the text classifier is based on $k$-nearest neighbors (kNN). 

\citet{evans2022} propose RAIDER, a system that can detect spear-phishing emails. To this purpose, a feature-evaluation system based on reinforcement learning has been implemented to select the best subset of features extracted from the email headers only. The authors found that RAIDER coupled with its automatic feature-selection process outperformed manual feature engineering. However, experiments were carried out only on public datasets (including the Enron corpus), which limits the scope of their results, as there are only about 44,000 emails.\\

Most of the presented techniques and systems primarily rely on features extracted from the email. Since a targeted attack is considered a rare event and an anomaly, it is common to leverage historical data, be it to assess the risk posed by a link or an email address, or to build an email sender profile~\cite{stringhini2015, duman2016, cidon2019, gascon2018, ho2017}. Note that historical data is not limited to emails. For instance, \citet{ho2017} derive features from historical user-login data and enterprise network-traffic logs. \\
tribution.

Natural-Language-Processing (NLP) techniques are applied in most of the presented approaches on textual content extracted from the email subject and body. \citet{stringhini2015} and \citet{duman2016} compute stylometric features (e.g., lexical and syntactic features) to determine the authorship of an email sender. However, they point out that the author identification can only be made with high accuracy if the textual content is long enough (at least 250 words~\cite{stringhini2015}). As most emails are short, they combine stylometric features with other features. While most systems presented in the literature extract features from the email textual content, a number of techniques rely primarily or solely on the content of email headers~\cite{ho2017, gascon2018, evans2022}. Table~\ref{tab:nlptechniques} summarizes some of the used NLP methods.% Extracted NLP features are mostly standard stylometric, syntactic, or structural features. Only \citet{han2016} proposes a semantic analysis of the email content with the detection of its topic.
\\

% Table ordered by year (and name)
\begin{table}[!ht]
    \centering
    \begin{tabular}{|c||C{10cm}|}
    \hline
        \textbf{Literature} & \textbf{Description} \\ \hline \hline
        \citet{stringhini2015} & Extraction of stylometric features (character occurrence, functional-word occurrence) \\ \hline
        \citet{duman2016} & Extraction of lexical features (number of words, character frequencies), syntactic features (part-of-speech tags such as adjectives and adverbs) and structural features (signature, contact information) \\ \hline
        \citet{han2016} & Extraction of layout features (subject and body text length), topic features (computed with Latent Semantic Indexing) and readability features (number of complex words, Fog index) \\ \hline
        \citet{gascon2018} & Extraction of most features from email headers and use of a bag-of-words model\\ \hline
        \citet{cidon2019} & Preprocessing of text (stop-word removal), and computation of a TF-IDF feature vector \\ \hline
        \citet{evans2022} & Extraction of features from email headers and use of a bag-of-words model\\ \hline
    \end{tabular}
    \caption{NLP techniques used in the literature.}
    \label{tab:nlptechniques}
\end{table}

Email headers also play a critical role in the literature. The \texttt{From} header is used to identify the sender of an email -- or the supposed sender in the case of an impersonation or email-account compromise. The recipient -- or recipients -- of an email can be determined with the \texttt{To}, \texttt{Cc}, and \texttt{Bcc} headers. \citet{gascon2018} compute transport features by analyzing the \texttt{Received} headers. These features are used to characterize the delivery path of an email, from its origin to its destination, as headers are added by email-transport agents along the way. \citet{han2016} also identify the source IP address, and derive features from it, such as the Autonomous System Number and the origin country. Transport features can however be poisoned with forged \texttt{Received} headers inserted by an attacker while composing the email. \citet{gascon2018} also derive composition features from email headers to characterize the particular email-client software used by a user. \citet{evans2022} do not extract a specific subset of relevant email headers, but consider email headers as raw features. These raw features are then evaluated and only the best ones are kept. \\

Some of the studied techniques aim at detecting URL-based targeted attacks, and as such extract features from the links found in the body of the email~\cite{stringhini2015, ho2017, cidon2019}. These features computed from links have two particularities. First, they rely on the collection of historical data at the user level (or organization level), to assess if the link or associated domain have already been seen in the context of the user (or in the context of the whole organization). Secondly, additional features can be derived from the links by leveraging particular databases or protocols (domain reputation with Alexa score, domain-registration date with WHOIS lookup, Autonomous System Number, and geographic location of IP address). \\

Regarding attachment-based targeted attacks, several systems do not extract any features from the files attached to the emails, as there are already existing technologies to detect malicious files (antivirus scanner, sandboxing)~\cite{ho2017, cidon2019}. In case attachment features are extracted, they are high-level features, such as the presence of an attachment, its type, extension, size, or associated fuzzy hash~\cite{stringhini2015, han2016, gascon2018}. \\

Table~\ref{tab:features} summarizes the different types of features used in the literature (headers, text, and links extracted from the email body, attachments, and historical data). \\

% Table ordered by year (and name)
\begin{table}[!ht]
    \centering
    \begin{tabular}{|c||c|c|c|c|c|}
    \hline
        \textbf{Literature} & \textbf{Headers} & \textbf{Text in body} & \textbf{Links in body} & \textbf{Attachments} &  \textbf{Historical data}\\ \hline \hline
        \citet{stringhini2015} & \checkmark & \checkmark & \checkmark & \checkmark & \checkmark\\ \hline
        \citet{duman2016} & \checkmark & \checkmark & ~ & ~ & \checkmark\\ \hline
        \citet{han2016} & \checkmark & \checkmark & ~ & \checkmark & ~ \\ \hline
        \citet{ho2017} & \checkmark & ~ & \checkmark & ~ & \checkmark\\ \hline
        \citet{gascon2018} & \checkmark & \checkmark & ~ & \checkmark & \checkmark \\ \hline
        \citet{cidon2019} & \checkmark & \checkmark & \checkmark & ~ & \checkmark\\ \hline
    \end{tabular}
    \caption{Features used in the literature.}
    \label{tab:features}
\end{table}

Table~\ref{tab:classificationalgorithms} presents the different classification algorithms used in the literature. Traditional machine-learning algorithms (random forest, SVM, kNN) are leveraged in most systems. \citet{cidon2019} justify the use of a random forest by its superior performance compared to decision trees, while still providing explainable inferences for analysis purposes. \citet{gascon2018} compare SVM and kNN, as SVM offers good performance with high-dimensional data, while kNN performs well with a handful of labeled targeted attacks. \citet{han2016} and \citet{ho2017} do not rely on classical machine-learning algorithms as these methods are likely to fail due to extreme class imbalance and lack of labeled targeted attacks. As an alternative, they propose novel techniques, such as semi-supervised learning model coupled with an affinity graph and Directed Anomaly Scoring. To the best of our knowledge, deep-learning techniques have not been explored, as they incur high computational cost both for training and inference~\cite{cidon2019}. 

% Table ordered by year (and name)
\begin{table}[!ht]
    \centering
    \begin{tabular}{|c||C{10cm}|}
    \hline
        \textbf{Literature} & \textbf{Algorithms} \\ \hline \hline
        \citet{stringhini2015} & Detection of sender impersonation with SVM \\ \hline
        \citet{duman2016} & Detection of sender impersonation with SVM \\ \hline
        \citet{han2016} &  Attribution of spear-phishing campaign with semi-supervised learning model coupled with an affinity graph \\ \hline
        \citet{ho2017} & Detection of URL-based targeted attacks with Directed Anomaly Scoring (DAS) \\ \hline
        \citet{gascon2018} & Detection of sender impersonation with SVM and kNN\\ \hline
        \citet{cidon2019} & Random forest for impersonation and link classifiers, kNN for text classifier \\ \hline
        \citet{evans2022} & Detection of sender impersonation with kNN \\ \hline
    \end{tabular}
    \caption{Classification algorithms used in the literature.}
    \label{tab:classificationalgorithms}
\end{table}

% Presentation of text-based attacks
\section{Text-Based Attacks}
\label{sec:text_attacks}

In this section, we focus on \textit{text-based attacks}, i.e., emails for which the fraudster primarily relies on the textual content to carry out their attack. As mentioned in Section~\ref{sec:contradictions}, the literature shows that the number of text-based targeted attacks has been raising recently. In the case of text-based attacks, the email potentially contains URLs or attachments, but those are not malicious by nature: it might be a URL that redirects to a benign website, or an attachment as a supporting document for the attack (e.g., invoice, banking details). Text-based attacks make up a significant part of fraud. Indeed, \citet{cidon2019} found that about 60\% of targeted attacks did not include any URL, while \citet{almusib2021} report that a URL or attachment is used in only 3\% of targeted attacks. We start by presenting several typologies of text-based non-targeted attacks.\footnote{\label{note1}All names or references to real entities and individuals have been changed in the following examples.} We then focus on different text-based targeted attacks.\cref{note1} During this study, we also analyze the social engineering techniques (manipulation principles and emotions) used by the fraudster -- techniques that were introduced in Section~\ref{sec:targeted_attacks}. Table~\ref{tab:textbasedattacks} summarizes the main findings for each typology of text-based attack: whether the attack is targeted or not, the associated goals, the social engineering techniques used, the individuals targeted within the organization (if relevant), and the different impersonation schemes found (impersonated individuals external to the organization are identified with an asterisk).

% Table ordered by year (and name)
\begin{table}[!ht]
    \centering
    \begin{tabular}{|c||c|c|C{2.4cm}|c|C{2.2cm}|C{2.2cm}|}
    \hline
        \textbf{Text-based attack} & \textbf{Targ.} & \textbf{Goals} & \textbf{Manipulation principles} & \textbf{Emotions} & \textbf{Targets} & \textbf{Who is impersonated?}\\ \hline \hline
        Bomb-threat scam & ~ & GC1 & Scarc., comm. & Fear & Anyone & ~ \\ \hline
        Sextortion & ~ & GC1 & Scarc., comm. & Fear & Anyone & ~ \\ \hline
        Advance-fee fraud & ~ & GC1 & Scarc., recip. & Greed & Anyone & ~ \\ \hline
        CEO fraud & \checkmark & GC1 & Auth., scarc., comm., liking & ~ & Finance dept. & CEO \\ \hline
        W-2 fraud & \checkmark & GC2 & Auth., scarc., comm., liking & ~ & HR, payroll dept. & Executive, tax accountant\text{*}, consultant\text{*} \\ \hline
        Gift-card scam & \checkmark & GC1 & Auth., scarc., comm., liking, recip. & Empathy & Office manager, exec. assistant, receptionist & Executive \\ \hline 
        Payroll fraud & \checkmark & GC1 & Scarc., comm., liking & ~ & HR, payroll dept. & Employee \\ \hline          
        Lawyer fraud & \checkmark & GC1, GC2 & Auth., scarc., comm., liking & ~ & Finance dept., executive & Lawyer\text{*}, executive \\ \hline
        Establish rapport & \checkmark & GC4 & Auth., scarc., comm., liking & ~ & Employee & Executive \\ \hline
    \end{tabular}
    \caption{Characteristics of text-based attacks.}
    \label{tab:textbasedattacks}
\end{table}

\subsection{Text-Based Non-Targeted Attacks}
As the name suggests, text-based non-targeted attacks do not target specific individuals, are generic, and are most often sent in bulk. A user should spot the fraudulent nature of these kinds of emails with ease as it appears most of the time out of context.

% Bomb-Threat Scam
% ----------------
% Goal: GC1
% Emotion: Fear
% Principle of manipulation: scarcity (because it's urgent), commitment and consistency (because people don't want others to be killed)
% Example: kreabs201812
\subsubsection{Bomb-Threat Scam}
In the case of a bomb-threat scam, the attacker claims that a bomb has been planted in the building where the victim is working, and that the bomb will be detonated unless a ransom is paid before a deadline. Some variations of this scam involve threats in public places, and the targets are usually city officials or law enforcement officers. The goal for the attacker is financial (GC1). 
Example 1 shows such a bomb-threat scam~\cite{krebs201812}. The attacker takes advantage of the victim's \textit{fear}, as the recipient is likely to be present in the building and be afraid to be killed or injured in case the bomb is detonated. Similarly, the recipient may also worry about relatives, friends, work colleagues, or acquaintances who may be present in or around the building. The attacker also relies on the \textit{scarcity} manipulation principle, as the situation has to be dealt with quickly (\say{You have to solve problems with the transfer by the end of the workday}, \say{If you are late with the money explosive will explode.}). The tone of the email and the vocabulary used leave no room for negotiation, which empowers the attacker and reinforces both the fear of the victim and the urgency of the situation. We can also note some elements of the \textit{commitment and consistency} principle. Indeed, the attacker assumes that the victim values the life of fellow humans (\say{in case of its detonation you will get many victims}).

\begin{Verbatim}
My man carried a bomb (Hexogen) into the building where your company is located. It is constructed under my direction. It can be hidden anywhere because of its small size, it is not able to damage the supporting building structure, but in the case of its detonation you will get many victims.
 
My mercenary keeps the building under the control. If he notices any unusual behavior or emergency he will blow up the bomb.
 
 
I can withdraw my mercenary if you pay. You pay me 20.000 $ in Bitcoin and the bomb will not explode, but don’t try to cheat -I warrant you that I will withdraw my mercenary only after 3 confirmations in blockchain network.
Here is my Bitcoin address : 1GHKDgQX7hqTM7mMmiiUvgihGMHtvNJqTv
 
 
You have to solve problems with the transfer by the end of the workday. If you are late with the money explosive will explode.
This is just a business, if you don’t send me the money and the explosive device detonates, other commercial enterprises will transfer me more money, because this isnt a one-time action.
 
 
I wont visit this email. I check my Bitcoin wallet every 35 min and after seeing the money I will order my recruited person to get away.
If the explosive device explodes and the authorities notice this letter:
We are not terrorists and dont assume any responsibility for explosions in other buildings.
\end{Verbatim}\label{ex:bombthreat}

% Sextortion
% ----------
% Goal: GC1
% Emotion: Fear 
% Manipulation principle: scarcity, commitment and consistency
% Example: krebs201807
\subsubsection{Sextortion}
In the context of sextortion, the attacker blackmails the victim for a financial gain (GC1). The attacker sends a message to the victim pretending to have hacked one of their devices and to be in the possession of compromising material. The attacker pretends in particular to have had access to the victim's webcam and to have recorded a video of the victim in the process of engaging in a sexual act. They threaten to forward the video to relatives and acquaintances unless the victim accepts to pay a ransom~\cite{paquet2019}. The ransom payment usually has to be made with cryptocurrencies (Bitcoin, Ethereum), and before a deadline. To convince the victim that a hack did happen, the attacker includes personal information about them, such as a password. The personal information was however acquired by other means, for example by purchasing on the dark web exposed data following a data breach (Yahoo!, Equifax, LinkedIn, Facebook, etc.).

Example 2 depicts a sextortion example \cite{krebs201807}. The attacker exploits the victim's \textit{fear}, and in particular the fear of being humiliated in the eyes of relatives, work colleagues, and acquaintances (\say{If I don’t get the payment, I will send your video to all of your contacts including relatives, coworkers, and so forth.}). The \textit{scarcity} manipulation principle is also used, as the victim needs to act quickly (\say{You have 24 hours in order to make the payment.}, \say{This is a non-negotiable offer, so don’t waste my time and yours by replying to this email.}). Finally, \textit{commitment and consistency} are also present, as the abuser assumes that the victim values their privacy and reputation -- including the reputation of their family.

\begin{Verbatim}
You don’t know me and you’re thinking why you received this e mail, right?
 
Well, I actually placed a malware on the porn website and guess what, you visited this web site to have fun (you know what I mean). While you were watching the video, your web browser acted as a RDP (Remote Desktop) and a keylogger which provided me access to your display screen and webcam. Right after that, my software gathered all your contacts from your Messenger, Facebook account, and email account.
 
What exactly did I do?
 
I made a split-screen video. First part recorded the video you were viewing (you’ve got a fine taste haha), and next part recorded your webcam (Yep! It’s you doing nasty things!).
 
What should you do?
 
Well, I believe, $1400 is a fair price for our little secret. You’ll make the payment via Bitcoin to the below address (if you don’t know this, search “how to buy bitcoin” in Google).
 
BTC Address: 1Dvd7Wb72JBTbAcfTrxSJCZZuf4tsT8V72
 
(It is cAsE sensitive, so copy and paste it)
 
Important:
 
You have 24 hours in order to make the payment. (I have an unique pixel within this email message, and right now I know that you have read this email). If I don’t get the payment, I will send your video to all of your contacts including relatives, coworkers, and so forth. Nonetheless, if I do get paid, I will erase the video immidiately. If you want evidence, reply with “Yes!” and I will send your video recording to your 5 friends. This is a non-negotiable offer, so don’t waste my time and yours by replying to this email.
\end{Verbatim}\label{ex:sextortion}

% Advance fee fraud
% -----------------
% Goal: GC1
% Emotion: Greed
% Manipulation principle: Scarcity, reciprocity
% Example: wein
\subsubsection{Advance-Fee Fraud}
When carrying out an advance-fee fraud, the attacker asks the victim to pay a certain sum of money while promising them to receive a significantly larger amount in return later on. The goal of the attacker is once again financial (GC1). If the victim gets tricked and does pay, the attacker either disappears or tries to convince them to execute additional payments. This type of fraud includes for example the traditional Nigerian scam also known as 419 fraud~\cite{holt2007}.

Example 3 illustrates this kind of fraud~\cite{wein}. We note two manipulation principles in this example. First, \textit{scarcity}, as the attacker offers a unique opportunity to the victim to earn a very large sum of money (\say{I am in position [sic] of Ten million U.S dollars (\$10,000,000.00)}, \say{we will be pleased to offer to you 15\% Of the total fund.}). Secondly, the fraudster exploits the \textit{reciprocity} principle. Indeed, they offer money to the victim in exchange for a service (\say{If you can be of sincere assistance to us}). Advance-fee fraud is an example of fraud that exploits the \textit{greed} of the victim, who is offered a once-in-a-lifetime financial opportunity.

\begin{Verbatim}
Subject: Miss.Blessing Jomo

Dearest One,

It is my pleasure to contact you for a business venture which I and my junior sister (Sandra) intend to establish in your country. Though I have not met with you before but I believe, one has to risk in other to succeed in life. I am in position of Ten million U.S dollars ($10,000,000.00) which my late Father deposited in security company Abidjan before he was assasinated by unknown persons during this war in Cote d'ivoire.

Now I and my sister want to move this money to invest it in your country or anywhere safe enough outside Africa for security and political reasons. We want you to help us claim and retrieve these fund from the company where it was deposited as family valuables and transfer it into your personal account in your country for investment purposes on these areas:
1). Telecommunication 
2). The Transport Industry
3). Five Star Hotel

If you can be of sincere assistance to us we will be pleased to offer to you 15% Of the total fund. I await your soonest response.

Respectfully yours,
Miss Blessing Jomo
\end{Verbatim}\label{ex:advancefee}

\subsection{Text-Based Targeted Attacks}
 Targeted attacks are more challenging to detect than non-targeted attacks, as already discussed in Section~\ref{sec:targeted_attacks}. Targeted attacks usually rely on the impersonation of a trusted entity, and mostly target specific individuals. Compared to URL-based targeted attacks and attachment-based targeted attacks, text-based targeted attacks come with additional difficulties. While the former can be detected with a URL scanner or an antivirus, text-based targeted attacks can only be detected by leveraging the email metadata, textual content, and -- if available -- historical data. Furthermore, the textual content is usually short, and does not appear out of context -- it could be easily mistaken with a legitimate email within the normal operation of the organization.

% CEO Fraud
% ---------
% Goal: GC1
% Who is impersonated: CEO or another executive
% Targeted: finance department
% Social engineering techniques: Authority, scarcity, commitment and consistency, liking
% Example: trustwave2016
\subsubsection{CEO Fraud}
The CEO-fraud attack is maybe the most famous \textit{corporate} text-based targeted attack. In this type of attack, the attacker impersonates the CEO, or another high-level executive of the company, and asks an employee -- generally someone from the finance department -- to carry out a financial action. For instance, the attacker may instruct the victim to perform a wire transfer to settle an outstanding invoice from a supplier. The email may directly include the payment details, or the attacker may provide these details after the reply of the victim to the first email. In some cases, the payment details are given in an attachment. The goal of the attacker is financial (GC1). CEO fraud usually contains urgency markers. They are often explicit in the textual content (\say{as soon as possible}, \say{asap}, \say{urgent}), and the message often appears to have been composed and sent from a smartphone (\say{Sent from my iPhone}) which implicitly reinforces the sense of urgency. It is also quite common that the CEO fraud is sent just before the end of the business day, or before the week-end. In this case, the victim is tired, lacks alertness, and is likely to expedite the task before leaving the office.

An example of CEO fraud is shown in Example 4~\cite{trustwave2016}. First, the attacker leverages the principle of \textit{authority} as the CEO is impersonated. It is likely that the employee performs the task requested by the main figure of authority within the company. Then, the attacker exploits the \textit{scarcity} principle, as the task has to be carried out before a deadline (\say{before the close of banking transactions for today}). This sense of urgency is also reinforced by the presence of an explicit marker in the subject line (\say{asap}). The attacker also uses the \textit{commitment and consistency} principle, as good employees value their work and want to make sure to keep up with the expectations of the company executives. Finally, the attacker exploits the \textit{liking} principle, as the CEO and the victim seem to know each other, as shown by the rather informal tone of the email, and as employees often want to obey their superior.

\begin{Verbatim}
Subject: Please get back to me asap.

Sue,

Please do you have a moment? Am tied up in a meeting and there is something I need you to take care of.
We have a pending invoice from our Vendor. I have asked them to email me a copy of the invoice. I will be highly appreciative if you can handle it before the close of banking transactions for today. I can't take calls now so an email will be fine.

Robert

\end{Verbatim}\label{ex:ceofraud}

% W-2 Fraud
% ---------
% Goal: GC2
% Who is impersonated: executive, external tax accounting company or individual
% Targeted: HR, payroll departments
% Social engineering techniques: Authority, scarcity, commitment and consistency, liking
% Example: barracuda2019
\subsubsection{W-2 Fraud}
The W-2 fraud is a corporate text-based targeted attack that affects exclusively US organizations. The goal of the attack is to collect the W-2 forms of the employees, which are confidential tax documents (GC2). This kind of form contains personal information of the employee (first name, last name, address), as well as the Social Security Number (SSN) that is a critical information, especially in the United States. The SSN can be used by the attacker to obtain the state and federal income tax refunds of the impersonated employee. The SSN may also be used to carry out identity theft -- for example to open a credit account. This fraud often targets the HR or payroll departments as they are the departments susceptible of holding those documents. The attacker usually impersonates an executive of the organization, or sometimes an external tax-accounting company or a consultant working for the organization. Urgency markers are common, especially in the context of tax filing where the tax return must be submitted before the tax-day deadline.

Example 5 presents a W-2 fraud \cite{barracuda2019}. The attacker first relies on the \textit{authority} principle, as an executive or figure of authority is impersonated, and the tone is directive (\say{I need}, \say{Kindly prepare}). The attacker also integrates \textit{scarcity} through an explicit urgency marker (\say{as soon as possible}). Once again, the attacker leverages the \textit{commitment and consistency} principle, as employees want to perform well at their job, and in particular meet the expectations of their managers. Finally, the \textit{liking} principle is also exploited, as the tone of the email shows that the impersonated executive and the victim know each other.

\begin{Verbatim}
Subject: 2018 W-2

Hi Jane,
I need W-2 copy list of all Employees wage and tax statement for 2018. Kindly prepare and attach the list in PDF file type and email them to me for review as soon as possible.

Thanks,

Daniel
\end{Verbatim}\label{ex:w2fraud}

% Gift card scam:
% Goal: GC1
% May be targeted or non-targeted
% Who is impersonated: executive
% Targeted: office manager, executive assistant, receptionist
% Social engineering technique: Authority, scarcity, commitment and consistency, reciprocity, liking
% Example: cofense2022
\subsubsection{Gift-Card Scam} \label{subsubsec:giftcardscam}
When carrying out a gift-card scam, the attacker requests the victim to purchase gift cards for a special occasion, such as a holiday or a birthday. It is usual that the attacker specifies the type of gift cards (Amazon, Target, Google Play, Apple, etc.), the amount (\$25, \$50, \$100), as well as the required quantity. The attacker may also instruct the victim to buy the gift card from a specific retail store (Walmart, Target, CVS, Walgreens, etc.). Once the gift cards are purchased, the attacker asks the victim to send the gift cards' numbers and PIN codes, either by typing them in or by sending a picture. The goal of the attacker is financial (GC1).

A gift-card scam may happen either in a non-targeted setting or in a targeted setting. In a non-targeted setting, these scams are sent in bulk, and the textual content is not personalized. Both enterprises and the general public are affected. In the context of an enterprise, this may involve the impersonation of an executive, with emails sent to many employees in a very short time frame. In a targeted setting, the scam relies on the impersonation of an executive, and the email is sent to the person in charge of purchasing gifts for the employees, such as the office manager, an executive assistant or a receptionist.

Examples 6a, 6b and 6c show the first, second, and third emails of a multi-stage gift-card scam~\cite{cofense2022}. In this attack, the fraudster first establishes a rapport with the target employee. This type of email is studied in Section~\ref{subsubsec:establishrapport}. It is very likely that the fraudster impersonates a figure of authority, as the targeted employee is willing to comply with his request. In a second email, the fraudster requests the employee to buy a gift card for a friend's daughter suffering from liver cancer. This is followed by a third email, where the fraudster is providing more specific details. We study these second and third emails in more details.

\begin{Verbatim}
Subject: Catch up

Hi there,

Just wondering if i can get a quick favour from you.

Thanks

Baily
\end{Verbatim}\label{ex:giftcardscam1}

In the second email (6b), the first manipulation principle is \textit{authority}, as it is very likely that the fraudster impersonates a figure of authority, and the tone supports this hypothesis (\say{I need you}, \say{let me know if you can handle this}). The second manipulation principle is \textit{scarcity}. Indeed, the fraudster pretends to be in an emergency situation and the tone is desperate (\say{It's her birthday today and I promised her to get it for her today, but I can't do this now}, \say{all my effort purchasing it online proved abortive}). Moreover, the fraudster relies on the principle of \textit{commitment and consistency} by claiming that the friend's daughter is suffering from cancer. The attacker assumes that the target employee is an empathetic person who values the well-being of others -- or has even experienced a similar tragedy and is willing to help a person in need. From an emotional point of view, the fraudster exploits the \textit{empathy} of the target employee. They also use the \textit{liking} principle as the impersonated person and the target employee seem to know each other. Finally, we can spot some elements of the \textit{reciprocity} principle, as the attacker promises to reimburse the target employee (\say{I'll reimburse you back as soon as possible.}). 

\begin{Verbatim}
Subject: Re: Catch up

Glad to hear from you.

I need you to get an Apple gift card for a friend daughter who is down with cancer of the Liver, it's her birthday today and I promised her to get it for her today, but I can't do this now because all my effort purchasing it online proved abortive. Can you get it from any store around you? I'll reimburse you back as soon as possible.

Kindly let me know if you can handle this.

Thanks
Baily
\end{Verbatim}\label{ex:giftcardscam2}

In the third email, the attacker indicates where to buy the cards (\say{You can get it from any shops like: Morrisons, Waitrose, Sainsbury’s, Argos, ASDA, John Lewis}), the type of cards desired (\say{Total amount needed is £200 in (£100 denominations)}), as well as the procedure for transmitting the desired information (\say{scratch off the back of the card to reveal the pin, then take a snapshot of the back showing the pin and have it email to me}. In this email, the attacker leverages the \textit{scarcity} principle as an urgency marker is present (\say{How soon can you get this done?}).

\begin{Verbatim}
Subject: Re: Catch up

Thank you so much. You can get it from any shops like: Morrisons, Waitrose, Sainsbury’s, Argos, ASDA, John Lewis. Total amount needed is £200 in (£100 denominations). I will need you to scratch off the back of the card to reveal the pin, then take a snapshot of the back showing the pin and have it email to me so I can forward it to her.

How soon can you get this done?

Thanks

Baily
\end{Verbatim}\label{ex:giftcardscam3}

% Payroll fraud:
% Goal: GC1
% Who is impersonated: employee
% Targeted: HR, payroll departments
% Social engineering technique: scarcity, commitment and consistency, liking
% Example: sslstore2021
\subsubsection{Payroll Fraud}
In the case of a payroll fraud, the attacker impersonates an employee and makes a request to HR or payroll departments to change the employee's direct-deposit information to a fraudulent bank account. The goal of the attacker is financial (GC1). The fraud relies on the sense of urgency. It is very common that the attacker explicitly expresses their wish that the next paycheck should be wired to the new bank account. In some cases, the attacker justifies the urgency by the fact that the previous bank account has been deactivated.

Example 7 illustrates this kind of fraud~\cite{sslstore2021}. First, the attacker exploits the \textit{scarcity} principle by creating an emergency situation (\say{I need your prompt assistance in this matter}). We can also point out the \textit{commitment and consistency} principle, as this attack targets an employee in the HR or payroll department who is -- presumably -- trying to provide quality services to other employees. Finally, the \textit{liking} principle is also leveraged, as it is likely that the impersonated employee and the targeted employee know each other, since the targeted employee works in the HR or payroll department.

\begin{Verbatim}
Subject: Payroll inquiry

Hello,

I have recently changed banks and need to change my direct deposit for payroll. I need your prompt assistance in this matter and when this will take effect.

Thank you.
\end{Verbatim}\label{ex:payrollscam}

% Lawyer fraud:
% Goal: GC1
% Who is impersonated: executive, lawyer, legal firm
% Targeted: financial department, other executive
% Social engineering technique: Authority, scarcity, commitment and consistency, liking
% Examples: abnormal2022, agari2020
\subsubsection{Lawyer Fraud}

Lawyer fraud is a corporate targeted attack based on the impersonation of a third-party lawyer. The attack also usually involves the impersonation of an executive of the targeted company. The targeted individual is typically an employee from the finance department, or another executive of the targeted company. The goal of the attacker is most of the time financial (GC1). In some cases, the goal is to obtain confidential information (GC2).

There are different variants of lawyer frauds. In a first variant, presented in Example 8a, the attacker impersonates a real lawyer from a well-known law firm (Clifford Chance, Sullivan \& Cromwell, Hogan Lovells, etc.), and sends a formal email to the targeted employee, requesting the payment of an overdue invoice~\cite{abnormal2022b}. The fraud relies on the \textit{authority} principle, as the employee may be intimidated by the formal aspect of the email, the reputation of the well-known law firm as well as the reputation of the impersonated lawyer. The \textit{scarcity} principle is also present as the invoice must be settled as soon as possible.
If the targeted employee does not respond positively, then the attacker impersonates an executive of the targeted company -- such as the CEO -- and sends another email to the victim. The fraud then relies again on the \textit{authority} principle, and encompasses the \textit{commitment and consistency} and \textit{liking} principles, as seen in other targeted attacks involving the impersonation of an executive.

\begin{Verbatim}
Subject: Unpaid Invoice

Good morning Sharon,

I am a lawyer from Sullivan & Cromwell LLP representing a client, chasing an unpaid invoice issued to your company. I have been advised to contact you on this matter and hoping you can get this settled as soon as possible

Kind Regards,

Whitney Chatterjee
Debt Collection Litigation Counsel
Sullivan & Cromwell LLP
1 New Fetter Ln, London EC4A 1AN
\end{Verbatim}\label{ex:lawyerfraud1}

In a second variant, the attacker impersonates the CEO of the targeted company, and sends a first email to another senior-level executive (Vice President, General Manager, Managing Director) requesting the targeted executive to work with an external legal counsel in the context of a business transaction or corporate expansion (e.g., M\&A) -- as shown in Example 8b \cite{agari2020}.  Later on, the attacker impersonates the external legal counsel and sends an email to the targeted executive, requiring a payment to finalize the transaction. This example of lawyer fraud relies on the \textit{authority} principle, as the impersonated CEO has the highest authority within the company, and the impersonated external legal counsel is perceived as an expert in the legal domain. The fraud also makes use of the \textit{scarcity} principle, as these operations follow a tight schedule and several urgency markers are present (\say{time-sensitive}, \say{by the close of this week}, \say{soonest}). The sense of urgency is implicitly reinforced by the fact that the email has been composed and sent with a smartphone (\say{Sent from my iPhone}). The \textit{commitment and consistency} and \textit{liking} principles are also exploited, as the impersonated CEO and targeted executive know each other and have a hierarchical relationship. The \textit{liking} principle is also strengthened by the fact the targeted executive is assigned to perform the task with trust and confidence (\say{I would like to entrust it to you}).

\begin{Verbatim}
Subject: Matter with law firm

Cheryl,

There is a time-sensitive corporate matter that needs to be resolved with our appointed law firm by the close of this week. I would like to entrust it to you.

Please let me know soonest by email if you can handle this and I will fill you in on the details.

Regards,
Bob

Sent from my iPhone
\end{Verbatim}\label{ex:lawyerfraud2}

% Establish Rapport:
% Goal: GC4
% Who is impersonated: executive
% Targeted: anyone in the organisation
% Social engineering technique: Build trust?
\subsubsection{Establish Rapport} \label{subsubsec:establishrapport}
It is quite common that a multi-stage targeted attack starts by one or several innocuous messages, where the attacker establishes a rapport with the target employee~\cite{cidon2019, talos2022}. The benefit is twofold for the attacker (GC4). First, they build trust with the target, who eventually lets their guard down. Second, they disarm certain email-security technologies that rely on communication history. The first message usually conveys a sense of urgency, and may also ask the target to provide their mobile-phone number to continue the conversation via text message or instant message (e.g., WhatsApp).

An illustration of establish rapport was introduced in Section~\ref{subsubsec:giftcardscam} (see Example 6a \cite{cofense2022}). Example 9 is another variant, where the goal of the attacker is to collect the phone number of the target employee \cite{talos2022}. The email first uses the \textit{authority} principle, as the CEO is impersonated (\say{Founder, CEO and Executive Chairman}), and the tone of the email is authoritative (\say{I have a task for you}). We also observe the \textit{scarcity} principle as urgency markers are present, both in the email subject line (\say{Emergency}) and the email body (\say{urgently}). As previously, the attacker additionally relies on the \textit{commitment and consistency} and \textit{liking} principles, due to the relationship between the impersonated CEO and the target employee.

\begin{Verbatim}
Subject: Emergency

Hello John

Hope you don't have so much on your plate? Well in case you do, please peg it because I have a task for you to carry out urgently.

Drop your number so i can brief you about it all.

Thanks
Sam
Founder, CEO and Executive Chairman at Awesome Inc.
\end{Verbatim}\label{ex:establishrapport}

% Conclusion
\section{Conclusion}
\label{sec:conclusion}

In this paper, we first highlighted that the literature does not propose clear and consistent definitions of spear phishing and Business Email Compromise (BEC) email-borne cyberattacks. However, it is crucial for researchers and security vendors to use a common vocabulary to develop techniques and products to tackle these threats. We thus introduced the notion of \textit{targeted attacks}, which can be fine-tuned by adding adjectives. We described the main characteristics of targeted attacks, such as their tailored nature, the type of attack vector, their rarity, and the social engineering techniques used by the attackers to trick the victims. We also studied the state of the art of targeted-attack-detection techniques. As a final contribution, we focused on text-based attacks, which pose a significant threat to organizations. We presented and compared text-based non-targeted attacks and text-based targeted attacks, with concrete examples and an in-depth analysis of the social engineering techniques used by the attacker.

\bibliographystyle{ACM-Reference-Format}
\bibliography{main}

%\appendix

\end{document}